\documentclass[twocolumn,prl]{revtex4}

\usepackage{graphicx}

\begin{document}

\date{\today}

\title{Universal Quantum Computation through Control of Spin-Orbit
Coupling}
\author{D. Stepanenko and N.E. Bonesteel}
\affiliation{National High Magnetic Field Laboratory and
Department of Physics, Florida State University, Tallahassee, FL
32310}

\begin{abstract}
We propose a method for quantum computation which uses control of
spin-orbit coupling in a linear array of single electron quantum
dots. Quantum gates are carried out by pulsing the exchange
interaction between neighboring electron spins, including the
anisotropic corrections due to spin-orbit coupling. Control over
these corrections, even if limited, is sufficient for universal
quantum computation over qubits encoded into pairs of electron
spins. The number of voltage pulses required to carry out either
single qubit rotations or controlled-Not gates scales as the
inverse of a dimensionless measure of the degree of control of
spin-orbit coupling.
\end{abstract}

\pacs{03.67.Lx, 73.21.La, 71.70.Ej}

\maketitle

Several quantum computation schemes are based on the idea of using
the spin-1/2 degrees of freedom of electrons or certain nuclei as
qubits \cite{loss98,kane98,vrijen00}.  In the proposal of Loss and
DiVincenzo \cite{loss98}, qubits are taken to be spins of single
electrons trapped in quantum dots. Two-qubit gates are then
carried out by switching on and off the exchange interaction
between neighboring spins \cite{burkard99, hu01}.

In the idealized limit of perfectly isotropic exchange, these
two-qubit gates conserve total spin and so have too much symmetry
to form a universal set --- i.e., they cannot be used to carry out
an arbitrary unitary transformation if qubits are represented by
single spins. A universal set can be realized if it is also
possible to perform arbitrary single spin rotations \cite{loss98},
but it is generally believed such rotations will be significantly
harder to achieve than two-qubit gates in the Loss-DiVincenzo
proposal. An attractive alternative is to use an encoding scheme
for which isotropic exchange alone is universal \cite{bacon00},
which requires encoding logical qubits into three or more physical
spins \cite{divincenzo00_2,hellberg03}.

Spin-orbit coupling leads to anisotropic corrections to the
exchange interaction \cite{kavokin0102} which, under certain
conditions elaborated on below, retains a residual rotational
symmetry about a fixed axis in spin space. For many purposes these
corrections are innocuous. The resulting exchange gates still form
a universal set when combined with single-spin rotations
\cite{burkard02,stepanenko03}. And, through a combination of pulse
shaping and locally defined spin quantization axes, they can be
made effectively isotropic, although in general only to second
order in spin-orbit coupling, so that exchange-only encoding
schemes can be used \cite{bonesteel01,wu03}.

The partial reduction in symmetry, from isotropic to axially
symmetric exchange, can also simplify the requirements for
universal quantum computation. For example, in \cite{kempe02} it
was shown that the XY interaction is universal for qubits encoded
into only two spins, provided there is a third ancillary spin for
each qubit. And in \cite{wu02} it was shown that any axially
symmetric anisotropic corrections, when combined with single spin
rotations about an axis perpendicular to the symmetry axis of the
exchange, can be used to construct a universal set of gates for
unencoded qubits.

In this Letter we propose a new method for quantum computation
based on the ability to {\it control} the spin-orbit induced
anisotropic corrections to the exchange interaction in a linear
array of quantum dots. Our proposal requires encoding logical
qubits into pairs of neighboring spins, similar to the encoding
used in \cite{levy02,benjamin01,lidar02}. However, unlike these
proposals, which require an inhomogeneous Zeeman field in addition
to exchange, our proposal employs only the spin-orbit corrected
exchange interaction.

The ${\bf k}$-dependent spin splitting of electronic bands due to
spin-orbit coupling is described by the Hamiltonian $H_{SO} = {\bf
\Omega}({\bf k})\cdot {\bf S}$ where ${\bf k}$ and ${\bf S}$ are,
respectively, the crystal momentum and spin of the electron.
Time-reversal symmetry requires that ${\bf \Omega}({\bf k}) = -
{\bf \Omega}(-{\bf k})$, thus ${\bf \Omega} \ne 0$ only in the
absence of inversion symmetry. For a [001] two-dimensional
electron gas (2DEG) in GaAs, or any other III-V zinc blende
semiconductor, there are two contributions to ${\bf \Omega}$.
Taking $k_x$ and $k_y$ to be along the [100] and [010] crystal
axes, respectively, the (linear) Dresselhaus contribution, ${\bf
\Omega}_D = f_D(-k_x,k_y,0)$, is due to  bulk inversion asymmetry,
with coupling $f_D$ inversely proportional to the square of the
width of the 2DEG \cite{dresselhaus55}, and the Rashba
contribution ${\bf \Omega}_R = f_R(k_y, -k_x,0)$ is due to the
structural inversion asymmetry of the quantum well used to form
the 2DEG \cite{rashba60}.

In the Hund-Mulliken description of two quantum dots, one Wannier
orbital is kept per dot. Let $t$ denote the tunneling amplitude
between these orbitals in the absence of spin-orbit coupling. The
effect of $H_{SO}$ is to induce a small spin precession during
this tunneling. If the dots lie in the [001] plane and are aligned
in the [110] direction, the precession axis is fixed to be along
the $[1\overline{1}0]$ direction \cite{cubic}. The precession
angle, $\eta$, then satisfies
\begin{eqnarray}
\tan \frac{\eta}{2} = s~\frac{a_0\omega_0}{\sqrt{2}t} \langle
\Psi_1| (k_x + k_y) | \Psi_2\rangle,
\end{eqnarray}
where $\Psi_i$ is the Wannier state associated with dot $i$, and
\begin{eqnarray}
s = \frac{f_D - f_R}{a_0\omega_0}
\end{eqnarray}
is a dimensionless measure of the strength of spin-orbit coupling.
Here $a_0$ and $\omega_0$ are, respectively, the linear size and
level spacing of a single isolated dot.

If the spin precession axis is fixed during gate operation, and
the $z$ axis in spin space is chosen to be parallel to this axis
(the [1${\overline 1}$0] direction in this case), exchange gates
in the presence of spin-orbit coupling will have the form
\cite{stepanenko03}
\begin{eqnarray}
U_{12}(\lambda; \alpha,\beta,\gamma) = e^{-i\lambda H},
\label{axial}
\end{eqnarray}
where
\begin{eqnarray}
H\! &=&\! {\bf S}_1 \cdot {\bf S}_{2} +\frac{\alpha}{2} (S_1^z -
S_{2}^z) +\beta(S_1^x S_{2}^y - S_1^y S_{2}^x)\nonumber\\
&&~~~~~~~~+\gamma(S_1^x S_{2}^x + S_1^y S_{2}^y)-\frac{1}{4}.
\label{axial2}
\end{eqnarray}
Here $\lambda$ is the integrated strength of the dominant
isotropic part of the interaction, and the parameters $\alpha$,
$\beta$ and $\gamma$ characterize deviations from perfect
isotropy.  The constant $-1/4$ in $H$ corresponds to a particular
choice for the overall phase of $U$ which will be convenient in
what follows. In \cite{stepanenko03} it was shown that for small
$s$, $\alpha = C_\alpha s$, $\beta = C_\beta s$ and $\gamma =
C_\gamma s^2$.  The dimensionless coefficients  $C_\beta$ and
$C_\gamma$ are both of order 1 and depend on the shape and
duration of the voltage pulse used to produce the gate, though
they cannot in general be set to 0. For a generic pulse,
$C_\alpha$ is also of order 1, but can be set to 0 by
time-symmetric pulsing.

We envision two methods for controlling these anisotropic
corrections. One is to control the width and shape of the
potential confining the electrons to the 2DEG, thus controlling
$f_D$ and $f_R$, and hence $s$. For the special case $f_D = f_R$
\cite{schliemann03}, $s$ can even be set to zero. The other is to
control the coefficients $C_{\alpha}$, $C_\beta$ and $C_\gamma$ by
pulse shaping, as described above (see also \cite{bonesteel01}).
Using these methods, it should be possible to achieve a {\it
continuous range} of gates of the form (\ref{axial}),
corresponding to small values of the parameter $s$. To ensure
approximate axial symmetry, we assume a linear array of $[001]$
quantum dots aligned along the $[110]$ direction, as shown in
Fig.~1. Note that corrections beyond Hund-Mulliken (i.e.,
involving more than one orbital per dot) will lead to deviations
from perfect axial symmetry and will be a source of error. Here we
assume these corrections are small enough to be ignored.

\begin{figure}[t]
\includegraphics[scale=.22]{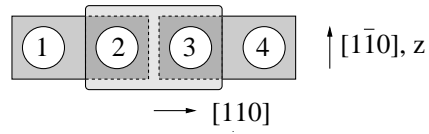} \vskip .15in
\caption{Four quantum dots forming two neighboring logical qubits,
12 and 34.  The dots lie in the [001] plane and are aligned along
the [110] direction.  The spin-orbit induced spin precession axis
is parallel to the [1$\overline{1}$0] direction. Exchange gates
between spins within a logical qubit are used for single qubit
rotations.  Two-qubit gates are carried out using exchange gates
acting on spins 2 and 3.} \label{4dots}
\end{figure}

Due to axial symmetry, the total $S_z$ quantum number of this
array will be conserved. It follows that the gates (\ref{axial})
cannot form a universal set if single spins are chosen to
represent qubits. We therefore adopt the two spin encoding scheme
of \cite{levy02,benjamin01,lidar02}.  To describe this encoding,
we associate a pseudospin space with every nearest-neighbor pair
of spins $i$ and $i+1$ spanned by the states
\begin{eqnarray}
|S\rangle_{i,i+1} =
\frac{1}{\sqrt{2}}(|\uparrow_i\downarrow_{i+1}\rangle
- |\downarrow_i\uparrow_{i+1}\rangle),\\
|T_0\rangle_{i,i+1} =
\frac{1}{\sqrt{2}}(|\uparrow_i\downarrow_{i+1}\rangle +
|\downarrow_i\uparrow_{i+1}\rangle).
\end{eqnarray}
where $|S\rangle_{i,i+1}$ is pseudospin up and
$|T_0\rangle_{i,i+1}$ is pseudospin down.  The Hilbert space
orthogonal to this pseudospin space is then spanned by the states
$|T_+\rangle_{i,i+1} = |\uparrow_i\uparrow_{i+1}\rangle$ and
$|T_-\rangle_{i,i+1} = |\downarrow_i\downarrow_{i+1}\rangle$.
Given our phase convention, the gates (\ref{axial}) leave this
space invariant,
\begin{eqnarray}
 U_{i,i+1}(\lambda,{\mbox{\boldmath$\phi$}})
|T_\pm\rangle_{i,i+1} = |T_\pm\rangle_{i,i+1},
\end{eqnarray}
and so are entirely determined by their action on the pseudospin
space,
\begin{eqnarray}
U_{i,i+1}(\lambda,{\mbox{\boldmath$\phi$}}) = e^{i\lambda/2} e^{-i
{{\mbox{\boldmath$\phi$}}}\cdot
{\mbox{\boldmath$\sigma$}}^{(i,i+1)}/2}. \label{rot}
\end{eqnarray}
Here $\mbox{\boldmath$\phi$} = \lambda(\alpha,\beta,\gamma+1)$ and
the components of ${\mbox{\boldmath{$\sigma$}}} =
(\sigma_x,\sigma_y,\sigma_z)$ are Pauli matrices, with the
superscript $(i,i+1)$ indicating that they act on the pseudospin
space associated with spins $i$ and $i+1$. These gates then
correspond to pseudospin rotations through the angle
\begin{eqnarray} \phi =
\lambda(1+2\gamma +\alpha^2 + \beta^2 + \gamma^2)^{1/2} =
\lambda+O(s^2), \label{mis}
\end{eqnarray}
about an axis parallel to ${\mbox{\boldmath$\phi$}}$.

\begin{figure}[t]
\includegraphics[scale=.25]{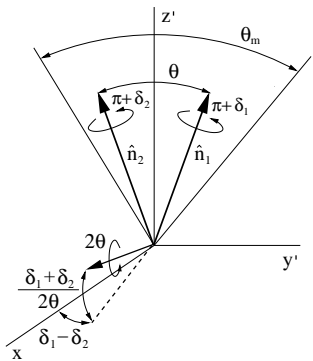} \vskip .15in
\caption{Rotation axes in the pseudospin space of two neighboring
spins.  The wedge lying in the plane perpendicular to $x$ and
sweeping out the angle $\theta_m$ contains rotation axes which can
be achieved using time-symmetric pulses and control of spin-orbit
coupling. Successive $\pi$ rotations about $\hat {\bf n}_1$ and
$\hat {\bf n}_2$, with $\hat {\bf n}_1 \cdot \hat {\bf n}_2 =
\cos\theta$, result in a $2\theta$ rotation about the $x$ axis.
The effect of errors in the rotation angles, $\delta_1$ and
$\delta_2$, on the net rotation axis is also shown. Here ${\hat
z}^\prime \parallel ( {\hat{\bf n}}_1 +\hat{\bf n}_2)$ and $\hat
y^\prime = {\hat z}^\prime \times {\hat x}$. }\label{rotation}
\end{figure}

In what follows we assume time-symmetric pulsing, so that $\alpha
= 0$ for all gates. The available pseudospin rotation axes will then
lie in the $yz$ plane. Allowing nonzero $\alpha$ through
time-asymmetric pulsing does not appreciably simplify any of our
constructions. Given the ability to control the remaining
anisotropic terms $\beta$ and $\gamma$, either through direct
control of $s$, or through pulse shaping, there will be a
continuous range of available rotation axes. For a given rotation
angle, $\phi$, these axes will sweep out a wedge shape in the $yz$
plane as shown in Fig.~\ref{rotation}. The degree of control of
spin-orbit coupling is then characterized by the angular size of
this wedge, which we denote $\theta_m$. We expect that $\theta_m$
will depend weakly on $\phi$, and will be on the order of the
largest possible value of $|s|$. Note that the wedge of allowed
rotation axes need not include the $z$ axis, corresponding to
$s=0$, although as noted above it may be possible to achieve this
through cancellation of the Dresselhaus and Rashba contributions.

For logical qubits encoded into the pseudospin spaces of dots $i$
and $i+1$, with $i$ odd, and computational basis states
$|0_L\rangle_{i,i+1} = |S\rangle_{i,i+1}$, and
$|1_L\rangle_{i,i+1} = |T_0\rangle_{i,i+1}$ (see Fig.~1), we now
show how pseudospin rotations can be used to perform single-qubit
rotations and controlled-NOT (CNOT) gates, thus providing a
universal set of quantum gates \cite{nielsen}.

Consider an arbitrary rotation about the $x$ axis. This operation
can be performed by a sequence of $\pi$ rotations about available
axes lying in the wedge. Figure~\ref{rotation} shows two such
axes, ${\bf n}_1$ and ${\bf n}_2$, making an angle $\theta \le
\theta_m$.  A $\pi$ rotation about ${\bf n}_1$  followed by a
$\pi$ rotation about ${\bf n}_2$  then results in a $2\theta$
rotation about the $x$ axis.  The sense of this rotation can be
reversed by reversing the order of the $\pi$ rotations. Since a
continuous range of axes within the wedge are available, a
rotation about the $x$ axis through an arbitrary angle $\Theta$
can be carried out by an even number, $2[\Theta/(2\theta_m)]+2$,
of $\pi$ rotations, where $[x]$ denotes the greatest integer
function of $x$.  The standard Euler construction can then be used
to generate arbitrary single qubit rotations, with the number of
pulses required growing as $1/\theta_m$ as $\theta_m$ goes to
zero.

As $\theta_m$ is reduced, this construction also becomes
increasingly sensitive to errors. To see this, let the rotation
angles about ${\bf n}_{1(2)}$ be $\pi +\delta_{1(2)}$ where
$\delta_{1(2)}$ are errors.  If we take the $z^\prime$ axis to be
parallel to ${\bf n}_1 + {\bf n_2}$ and the $y^\prime$ axis
parallel to $\hat z^\prime \times \hat x$ then the composition of
these two rotations will yield an overall $2\theta +
O(\delta^2/\theta)$ rotation about an axis deviating from the
$\hat x$ axis by an angle $\delta_1 - \delta_2$ in the $y^\prime$
direction and $(\delta_1 + \delta_2)/2\theta$ in $z^\prime$
direction  (see Fig.~\ref{rotation}). Thus, the larger $\theta_m$
is, the more robust this construction is against errors.

Now consider the two logical qubits shown in Fig.~1. A two-qubit
gate between the 12 qubit and the 34 qubit can be carried out by a
sequence of pulses acting on spins 2 and 3. Because the pseudospin
space of spins 2 and 3 does not correspond to a logical qubit,
rotations in this space will, in general, mix in noncomputational
states resulting in leakage errors. To avoid such errors, the net
unitary transformation must be diagonal in the $\{\uparrow_1
\downarrow_2 \uparrow_3 \downarrow_4$, $\uparrow_1 \downarrow_2
\downarrow_3 \uparrow_4$, $\downarrow_1 \uparrow_2 \uparrow_3
\downarrow_4$, $\downarrow_1 \uparrow_2 \downarrow_3 \uparrow_4\}$
basis of the four spins. The most general unitary operator of the
form (\ref{rot}) for which this is the case consists of a rotation
about the $x$-axis in pseudospin space. It follows that the net
gate must be of the form
\begin{eqnarray}
U_{23}(\Lambda,\Phi) = \prod_k
U_{23}(\lambda_k;{\mbox{\boldmath{$\phi$}}}_k) = e^{i\frac{\Lambda}{2}}
e^{-i \frac{\Phi}{2} \sigma^{(2,3)}_x}, \label{U}
\end{eqnarray}
where $\Lambda = \sum_k \lambda_k$ is the net phase and $\Phi$ is
the rotation angle about the $x$-axis produced by the sequence of
rotations $\{{\mbox{\boldmath{$\phi$}}}_k\}$.   Note that both
$\Lambda$ and $\Phi$ are defined modulo $4\pi$.

The gate (\ref{U}) can be expressed in terms of operators acting
on the logical qubits as follows,
\begin{eqnarray}
U_{23}(\Lambda,\Phi) = e^{i\frac{\Lambda}{4}}
e^{i\frac{\Lambda}{4} \sigma^{(1,2)}_x\sigma^{(3,4)}_x}
e^{i\frac{\Phi}{4}\sigma_x^{(1,2)}}
e^{i\frac{\Phi}{4}\sigma_x^{(3,4)}}. \label{nearlycanonical}
\end{eqnarray}
By casting this gate in its canonical form \cite{kraus01}, it can
be shown to be equivalent to a CNOT gate, up to single qubit
rotations, if and only if
\begin{eqnarray}
\Lambda = \sum_k \lambda_k = (2n+1)\pi. \label{constraint}
\end{eqnarray}
Below we outline two procedures for simultaneously satisfying
(\ref{U}) and (\ref{constraint}).

For the first procedure, let $R_x(\pi)$ be a $\pi$ rotation about
the $x$ axis. Using the single qubit rotation scheme described
above, this rotation can be performed through a sequence of $2n =
2[\pi/(2\theta_m)]+2$ rotations about available axes. If $A(\phi)$
is then a $\phi$ rotation about a particular available axis lying
in the $yz$ plane,  the sequence of rotations $A(\phi) R_x(\pi)
A(\phi)$ will have the form (\ref{U}) with $\Phi=(2n+1)\pi$
regardless of the value of $\phi$. According to (\ref{mis}) the
contribution of $R_x(\pi)$ to the total phase $\Lambda$ will then
be $2n\pi + \mu$ where $\mu \sim O({s^2/\theta_m}) \sim O(s)$. To
satisfy (\ref{constraint}) we therefore require $\phi = \pi/2 +
O(s)$, where the $O(s)$ adjustment must be chosen so that $\lambda
= \pi/2 - \mu/2$ for $A(\phi)$ and thus $\Lambda = (2n+1)\pi$.
This procedure is similar to those proposed in the two spin
encoding schemes of \cite{benjamin01,levy02,lidar02,wu02}. The
main difference is that in these constructions the $R_x$ rotation
is generated by an inhomogeneous Zeeman field, whereas in ours it
is generated entirely by a sequence of exchange gates
corresponding to $\pi$ rotations in the wedge of available axes.
As for single qubit rotations, as $\theta_m$ goes to zero, the
number of required pulses scales as $1/\theta_m$ and the
construction becomes increasingly sensitive to errors.

The second procedure requires more pulses in the limit of small
$\theta_m$, but is simpler and less susceptible to error. The idea
is to perform a sequence of $2\pi$ pseudospin rotations about any
available axis or axes and use the spin-orbit induced mismatch
between $\phi$ and $\lambda$ to accrue the extra $\pi$ phase
required to satisfy (\ref{constraint}). The resulting gate will
then have the form (\ref{U}) with $\Phi = 2n\pi$ where $n$ is the
number of $2\pi$ rotations. According to (\ref{mis}), for the
$i$th rotation the corresponding phase factor will be $\lambda_i =
2\pi + \nu_i$, where $\nu_i \sim O(s^2)$. For a sequence to
satisfy the constraint (\ref{constraint}) the sum of all phases,
and hence $\sum_i \nu_i$,  must be an {\it odd} multiple of $\pi$.
Given control of spin-orbit coupling, there will be a continuous
range of achievable $\nu$ values for each 2$\pi$ rotation, with
$\nu_1 < \nu < \nu_2$ where $\nu_1,\nu_2 \sim O(s^2)$.  If this
range includes $0$ then (\ref{constraint}) can always be satisfied
with $[\pi/\nu_{max}]+1$ rotations, where $\nu_{max} = {\rm
max}(|\nu_1|,|\nu_2|)$.  If this range does not include $0$ it
will still always be possible to satisfy (\ref{constraint}) with,
at most, $[\nu_{max}/(\nu_2 - \nu_1)]+2+[\pi/\nu_{max}]$
rotations.

\begin{figure}[t]
\includegraphics[scale=0.11]{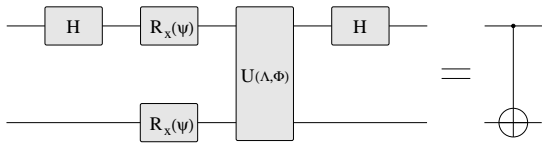}\vskip .15in
\caption{Proposed CNOT construction.  Each line corresponds to a
logical qubit.  $U(\Lambda,\Phi)$ is defined in (\ref{U}) with
$\Lambda = (2n+1)\pi$.  The value of $\Phi$ depends on the
procedure used to carry out the CNOT. $H=(\sigma_x +
\sigma_z)/\sqrt{2}$ is a Hadamard gate and $R_x(\psi)$ is a single
qubit rotation about the $x$ axis through an angle $\psi$ equal
modulo $2\pi$ to $(\Phi + \Lambda)/2$.} \label{system}
\end{figure}

Regardless of which procedure is used, single qubit gates acting
on logical qubits $12$ and $34$ are required to complete the CNOT
construction.  One procedure for doing this is shown in Fig.~3.

Initialization can be performed by switching on the exchange
interaction between pairs of spins forming logical qubits and
cooling. If $s$ can be set to 0 for this initialization, the
logical qubits will equilibrate to the $|0_L\rangle$ state. If $s$
cannot be set to $0$, logical qubits will still equilibrate to a
particular state which can then be rotated to $|0_L\rangle$ by
single qubit rotations. Readout of logical qubits can be performed
using a modified version of the measurement scheme proposed by
Kane \cite{kane98}. By switching on the tunneling between dots
forming a logical qubit, and raising the voltage of one dot so
that it will become doubly occupied if and only if the final state
is a singlet, the qubit measurement can be converted to a charge
measurement which can be performed using a single electron
transistor. As for initialization, if the spin-orbit induced spin
precession cannot be turned off during this process, it will not
correspond to a measurement in the $\{|0_L\rangle$,$|1_L\rangle\}$
basis, but rather a measurement along a pseudospin axis which is
nearly parallel to $z$.  Again this does not cause any fundamental
problems.

To summarize, we propose a method for quantum computation which
exploits the ability to control the anisotropic corrections to the
exchange interaction due to spin-orbit coupling. The degree of
control of these corrections is characterized by the parameter
$\theta_m$. For two spin encoding of logical qubits, this control
can be used to carry out single logical qubit rotations and CNOT
gates, with the number of pulses required for each scaling as
$1/\theta_m$ in the limit of small $\theta_m$.  For this scheme to
be useful, it is clearly desirable to design a system for which
$\theta_m$ is as large as possible.

This work is supported by the National Science Foundation through
NIRT Grant No.\ DMR-0103034.


\begin{references}

\bibitem{loss98} D. Loss and D.P. DiVincenzo, Phys. Rev. A {\bf 57},
120 (1998).

\bibitem{kane98} B.E. Kane, Nature {\bf 393}, 133 (1998).

\bibitem{vrijen00} R. Vrijen et al., Phys. Rev. A {\bf 62}, 012306
(2000).

\bibitem{burkard99} G. Burkard, D. Loss, and D.P. DiVincenzo,
Phys. Rev. B {\bf 59}, 2070 (1999).



\bibitem{hu01} X. Hu and S. Das Sarma, Phys. Rev.
A {\bf 61}, 062301 (2000).

\bibitem{bacon00} D. Bacon et al., Phys. Rev. Lett. {\bf 85}, 1758 (2000).

\bibitem{divincenzo00_2} D.P. DiVincenzo et al., Nature {\bf 408}, 339 (2000).

\bibitem{hellberg03} C.S. Hellberg, Preprint. quant-ph/0304150.

\bibitem{kavokin0102} K.V. Kavokin, Phys. Rev. B {\bf 64}, 075305
(2001), K.V. Kavokin, Phys. Rev. B {\bf 69}, 075302 (2004).

\bibitem{burkard02} G. Burkard and D. Loss, Phys. Rev. Lett. {\bf
88}, 047903 (2002).

\bibitem{stepanenko03} D. Stepanenko et al., Phys. Rev. B {\bf
68}, 115306 (2003).


\bibitem{bonesteel01} N.E. Bonesteel, D. Stepanenko, and
D.P. DiVincenzo, Phys. Rev. Lett. {\bf 87}, 207901 (2001).

\bibitem{wu03} L.-A. Wu and D.A. Lidar, Phys. Rev. Lett. {\bf 91},
097904 (2003).


\bibitem{kempe02} J. Kempe and K.B. Whaley, Phys. Rev. A {\bf 65},
052330 (2002).

\bibitem{wu02} L.-A. Wu and D.A. Lidar, Phys. Rev. A {\bf 66}, 062314
(2002).

\bibitem{levy02} J. Levy, Phys. Rev. Lett. {\bf 89}, 147902 (2002).

\bibitem{benjamin01} S.C. Benjamin, Phys. Rev. A {\bf 64}, 054303
(2001).

\bibitem{lidar02} D.A. Lidar and L.-A. Wu, Phys. Rev. Lett. {\bf 88},
017905 (2002).

\bibitem{dresselhaus55} G. Dresselhaus, Phys. Rev. {\bf 100}, 580
(1955); M.I. Dyakonov and V.Yu. Kachorovskii, Sov. Phys.
Semicond. {\bf 20}, 110 (1986).


\bibitem{rashba60} E.I. Rashba, Fiz. Tv. Tela (Leningrad) {\bf 2},
1224 (1960) [Sov. Phys. Solid State {\bf 2}, 1109 (1960)]; Y.A.
Bychkov and E.I. Rashba, J. Phys. C {\bf 17}, 6039 (1984).




\bibitem{cubic}  Here we assume symmetry of the in-plane fields with
respect to reflection through the [1${\overline 1}$0] plane. Note
also that including the cubic contribution to ${\bf \Omega}_D$
does not alter our conclusion that the precession axis is fixed to
be parallel to [1${\overline 1}$0] for our dot geometry.


\bibitem{schliemann03} John Schliemann, J.C. Egues and Daniel
Loss, Phys. Rev. Lett. {\bf 90}, 146801 (2003).



\bibitem{nielsen} M.A. Nielsen and I.L. Chuang, {\it Quantum
Computation and Quantum Information} (Cambridge University Press,
Cambridge, UK, 2000).



\bibitem{kraus01}B. Kraus and J.I. Cirac, Phys. Rev. A {\bf 63}, 062309
(2001); K. Hammerer, G. Vidal, J.I. Cirac, Phys. Rev. A {\bf 66},
062321 (2002)


\end{references}
\end{document}